\documentclass[%
 reprint,
 amsmath,amssymb,
 aps,
]{revtex4-1}
\usepackage{color}
\usepackage{array,multirow} 
\usepackage{graphicx}
\usepackage{dcolumn}
\usepackage{bm}
\usepackage[utf8]{inputenc}
\begin{document}

\preprint{APS/123-QED}

\title{ Linking axionlike dark matter to neutrino masses }

\author{C. D. R. Carvajal}
\email{cdavid.ruiz@udea.edu.co}
\affiliation{Instituto de F\'isica, Universidad de Antioquia,\\ Calle 70 No. 52-21, Medell\'in, Colombia}%
\affiliation{INFN, Laboratori Nazionali di Frascati,\\ C.P. 13, 100044 Frascati, Italy}

\author{B. L. S\'anchez-Vega}
\email{brucesan@ift.unesp.br}
\affiliation{Universidade Estadual Paulista (Unesp),  Instituto de F\'isica Te\'orica (IFT), S\~ao Paulo. R. Dr. Bento Teobaldo Ferraz 271, Barra Funda, S\~ao Paulo - SP, 01140-070, Brasil}

\author{O. Zapata}
\email{oalberto.zapata@udea.edu.co}
\affiliation{Instituto de F\'isica, Universidad de Antioquia,\\ Calle 70 No. 52-21, Medell\'in, Colombia}%

\date{\today}
          
\begin{abstract}
We present a framework linking axionlike particles (ALPs) to neutrino masses through the minimal inverse seesaw (ISS) mechanism in order to explain the dark matter (DM) puzzle. Specifically, we explore three minimal ISS cases where mass scales are generated through gravity-induced operators involving a scalar field hosting ALPs. In all of these cases, we find gravity-stable models providing the observed DM relic density and, simultaneously, consistent with the phenomenology of neutrinos and ALPs. Remarkably, in one of the ISS cases, the DM can be made of ALPs and sterile neutrinos. Furthermore, other considered ISS cases have ALPs with parameters inside regions to be explored by proposed ALPs experiments.
\end{abstract}

\maketitle


\section{Introduction\label{introduction}}
The discovery of neutrino oscillations \cite{Fukuda:1998mi,Ahmad:2002jz} and the fact that baryonic matter only yields a few percent contribution to the energy density of the Universe \cite{Ade:2015xua} are two experimental evidences calling for physics beyond the standard model (SM). On the theoretical side, the apparent absence of CP violation in the QCD sector is also a strong motivation for going beyond the SM since it can be dynamically explained by the Peccei-Quinn mechanism \cite{Peccei:1977hh}, which requires to extend the SM gauge group with a global symmetry and the existence of a pseudo-Nambu-Goldstone boson, the axion \cite{Weinberg:1977ma,Wilczek:1977pj}. 
Besides elegantly solving the strong CP problem \cite{Kim:1979if,Shifman:1979if,Dine:1981rt,Zhitnitsky:1980tq}, the Peccei-Quinn mechanism may be also related to the solution of DM and neutrino puzzles by offering a candidate for cold DM, the axion itself, \cite{Preskill:1982cy,Abbott:1982af,Dine:1982ah} and a connection to the neutrino mass generation \cite{Mohapatra:1982tc,Shafi:1984ek,Langacker:1986rj,Shin:1987xc,He:1988dm,Berezhiani:1989fp,Bertolini:1990vz}. 

In the same vein, ALPs, arising from spontaneous breaking of approximate global symmetries, are also theoretically well motivated since these appear in a variety of ultraviolet extensions of the SM \cite{Jaeckel:2010ni,Ringwald:2012hr,Dias:2014osa,Montero:2011tg} and, as in QCD axion models, these can make up all of Universe DM \cite{Arias:2012az}, or be a portal connecting the DM particle to the SM sector \cite{Nomura:2008ru}. Moreover, there are some astrophysical phenomena such as the cosmic $\gamma$-ray transparency \cite{DeAngelis:2007dqd,Simet:2007sa,SanchezConde:2009wu,Meyer:2013pny}, the x-ray excess from the Coma cluster \cite{Conlon:2013txa,Angus:2013sua} and the x-ray line at 3.55 keV \cite{Bulbul:2014sua,Boyarsky:2014jta} that suggest the presence of ALPs. These hints have led to a plethora of search strategies involving astrophysical observation production and detection in laboratory experiments \cite{Jaeckel:2010ni,Ringwald:2012hr,Essig:2013lka} with the aim of establishing the ALPs properties. 

In the context of ALPs models, the approximate continuous symmetry is typically assumed to be remnant of an exact discrete gauge symmetry as gravity presumably breaks the global symmetries through Planck-scale suppressed operators. In other words, since the global symmetry is highly unstable it is usually stabilized by imposing a discrete gauge symmetry \cite{Ibanez1991291,PhysRevD.45.1424,Ibanez1993301,Luhn:2008sa} such as a $Z_N$ symmetry \cite{Dias:2014osa,Carvajal:2015dxa,Kim:2015yna} (see Refs.  \cite{Dine:1992vx,Dias:2002hz,Dias:2002gg,Babu:2002ic,Dias:2003zt,Carpenter:2009zs,Harigaya:2013vja,Ringwald:2015dsf} for $Z_N$ realizations in QCD axion models).

This discrete gauge symmetry protects the ALPs mass against large gravity-induced corrections and it can also be used to stabilize other mass scales present in the theory. In particular, with the aim of generating neutrino masses the authors in Refs. \cite{Dias:2014osa,Carvajal:2015dxa} used these types of discrete gauge symmetries in order to protect the associated lepton-number-breaking scale. In this work, we go further by building a self-consistent framework of ALPs DM\footnote{Note that in \cite{Dias:2014osa,Carvajal:2015dxa} the ALPs is used to explain some astrophysical anomalies and not to give account for the entire DM abundance.} and neutrino masses via the ISS mechanism \cite{Mohapatra:1986aw,Mohapatra:1986bd}. For this purpose we make use of appropriate $Z_N$ discrete gauge symmetries to protect the suitable ALPs mass reproducing the correct DM relic abundance as well as  to stabilize the mass scales present in the ISS mechanism. It turns out that the ISS mass terms are determined -up to some factor- by $v^n_\sigma/M^{n-1}_{\textrm{Pl}}$, where $v_\sigma$ is the vacuum expectation value (VEV) of the scalar field $\sigma$ that spontaneously breaks the global symmetry U$(1)_A$ and hosts the ALPs, $a$. $n$ is an integer that is determined by the invariance of such terms under the symmetries of the model and some phenomenological constraints.   
 
In order to implement the ISS mechanism we extend the SM matter content by introducing $n_{N_{R}} (n_{S_{R}})$ generations of SM-singlet fermions $N_{\textrm{R}} (S_{\textrm{R}})$ as it is usual. In this work, we consider the minimal number of singlet fermionic fields that allows to fit all the experimental neutrino physics: the $(n_{N_{R}}, n_{S_{R}})=(2,2),$ $(2,3)$ and $(3,3)$ cases  \cite{Abada:2014vea}. In each case, the ALPs plays the role of the DM candidate. Moreover, for the (2,3) ISS case there is a possibility of having a second DM candidate: the sterile neutrino (the unpaired singlet fermion) \cite{Kusenko:2009up,Abazajian:2012ys,Adhikari:2016bei}.  
Motivated by that, we also build a multicomponent DM framework where the DM of the Universe is composed by ALPs and sterile neutrinos, with the latter being generated through the active-sterile neutrino mixing \cite{Dodelson:1993je} and accounting for a fraction of the the DM relic density.

As far as phenomenological issues are concerned, since in each framework the  approximate continuous symmetry is anomalous respect to the electromagnetic gauge group (through an exotic vectorlike fermion) instead of being anomalous respect to QCD, it is possible to build an effective interaction term involving the axion field and the electromagnetic field strength and its dual \cite{Srednicki:1985xd,Kaplan:1985dv}. This in turn implies that the ALPs may be detected in current- and/or proposed- experiments that use the ALPs-photon coupling as their main interaction channel to search for ALPs \cite{Jaeckel:2010ni,Ringwald:2012hr,Essig:2013lka,Graham:2015ouw}. Moreover, considering this particle as the dark matter candidate it can be part of the Milky Way DM halo and could resonantly convert into a monochromatic microwave signal in a microwave cavity permeated by a strong magnetic field \cite{Sikivie:1983ip,Sikivie:1985yu,Sikivie:2006ni}. On the other hand, since a large portion of the parameter space of ALPs (e.g. low masses and couplings) is relatively unconstrained by experiment since the conventional experiments, Helioscopes, Haloscopes and others \cite{Graham:2015ouw}, are only sensitive to axion particles whose Compton wavelength is comparable to the size of the resonant cavity, it is important looking for new search strategies in order to cover other regions of the parameter space. To reach smaller values for the ALPs mass and ALPs-photon coupling is necessary a different experimental approach  like the ones associated to the ABRACADABRA proposal \cite{Kahn:2016aff,Sikivie:2013laa},  where it is suggested a new set of experiments based on either broadband or resonant detection of an oscillating magnetic flux,  designed for the axion detection in the range $m_a\in[10^{-14},10^{-6}]$ eV. And it is precisely these kinds of searches that can be used to probed the benchmark regions that we study within the (2,2) and (3,3) ISS cases. 

The rest of the paper is organized as follows: in Sec. \ref{framework}  we discuss phenomenological and theoretical conditions that lead to a successful protection of the ALPs mass and the ISS texture against gravity effects. In Sec. \ref{model} we search for viable models simultaneously compatible with DM phenomenology, neutrino oscillation observables and lepton-flavor-violating processes. Finally, we present our discussion and conclusions in Sec \ref{conclusion}.

\section{Framework \label{framework}}
The goal of this section is to present the main ingredients of a SM
extension in order to link ALPs to neutrino
mass generation, and at the same time, to offer an explanation for
the current DM relic density reported by Planck Collaboration \cite{Ade:2015xua}. In
order to achieve that, the SM matter content must to be extended with
some extra fields. Besides the scalar $\sigma$ and fermionic  $S_{\textrm{R}\alpha}$ and $N_{\textrm{R}\beta}$ fields, an extra electrically charged fermion $E$ is also
added to the SM to make possible the coupling of ALPs to photons, $g_{a\gamma}$. That is necessary because $g_{a\gamma}$ is anomaly induced and there is no any U$\left(1\right)_{A}$ symmetry anomalous in the electromagnetic group just with the SM charged fermions.  The main role of the anomalous U$(1)_A$ symmetry is to induce an ALPs coupling to two photons. This brings as a consequence that the ALPs can be found, in principle, in current and/or proposed experiments that make use of the ALPs-photon coupling. We will show target regions of some experiments searching for ALPs in Fig \ref{fig1}. Also, it will be found that ALPs in some ISS cases discussed in this paper are inside the regions of planned experiments \cite{Kahn:2016aff,Sikivie:2013laa}.

Another key point of the framework is the existence of a $Z_{N}$
discrete gauge symmetry. In order to understand its role, firstly,
note that to impose an anomalous U$\left(1\right)_{A}$ symmetry to
the Lagrangian does not seem sensible in the sense that in the absence
of further constraints on very high energy physics we should expect
all relevant and marginally relevant operators that are forbidden
only by this symmetry to appear in the effective Lagrangian with coefficient
of order one. However, if this symmetry follows from some other free
anomaly symmetry, in our case from the a $Z_{N}$ discrete gauge symmetry,
all terms which violate it are then irrelevant in the renormalization
group sense. Secondly, the $Z_{N}$ symmetry also protects both the
ALPs mass and the ISS texture against gravity effects as we will explain
in more detail later on. For these reasons, the effective Lagrangian
will be invariant under a $Z_{N}$ discrete gauge symmetry. Due to the ALPs mass is very low and only protected by the U$(1)_A$ symmetry which is explicitly broken by gravity effects, the Z$_N$ symmetry will have a high order. This fact also happens in models with QCD axions and it is shared by all models with this type of stabilization mechanism \cite{PhysRevLett.118.031801,Dias:2014osa,Ringwald:2015dsf,Dias:2002hz,Dias:2002gg,Carvajal:2015dxa}.

\subsection{Lagrangian \label{sec21}}

The effective Lagrangian that we consider to relate the
ISS mechanism to ALPs DM reads
\begin{equation}
\mathcal{L}\supset\mathcal{L}_{\text{SM}}^{\text{Yuk}}+\mathcal{L}_{\sigma}+\mathcal{L}_{\textrm{ISS}}+\mathcal{L}_{E},\label{eq1-1}
\end{equation}
where $\mathcal{L}_{\text{SM}}^{\text{Yuk}}$ is nothing more than
the Yukawa Lagrangian of the SM
\begin{equation}
\mathcal{L}_{\text{SM}}^{\text{Yuk}}=Y_{ij}^{(u)}\overline{Q_{\textrm{L}i}}\widetilde{H}u_{\textrm{R}j}+Y_{ij}^{(d)}\overline{Q_{\textrm{L}i}}Hd_{\textrm{R}j}+Y_{ij}^{(l)}\overline{L_{i}}Hl_{\textrm{R}j}+\text{H.c.},\label{eq2-1}
\end{equation}
with the usual $Q_{\textrm{L}i},u_{\textrm{R}i},d_{\textrm{R}i}$
and $L_{i},l_{\textrm{R}i}$ fields denoting the quarks and leptons
of the SM, respectively. $H$ is the Higgs SU$(2)_{L}$ doublet with
$\widetilde{H}=i\tau_{2}H^{*}$ ($\tau_{2}$ is the second Pauli matrix). 

The term in $\mathcal{L}_{\sigma}$ (Lagrangian involving the $\sigma$
field) which is relevant in our discussion is the following non-renormalizable operators
\begin{equation}
\mathcal{L}_{\sigma}\supset g\frac{\sigma^{D}}{M_{\textrm{Pl}}^{D-4}}+\text{H.c.},\label{eq3-1}
\end{equation}
with $g=e^{i\delta}\left|g\right|$ and $D$ being an integer. The $\sigma$ field
 is parametrized as $\sigma(x)=\frac{1}{\sqrt{2}}[v_{\sigma}+\rho(x)]e^{i\frac{a(x)}{v_{\sigma}}},$
with $a(x)$ being the ALPs field and $\rho(x)$ the radial part that
will gain a mass of order of the vacuum expectation value \cite{Ringwald:2014vqa,Dias:2014osa,Marsh:2015xka}
\begin{equation}
10^{9}\lesssim\sqrt{2}\left\langle \sigma\right\rangle \equiv v_{\sigma}\lesssim10^{14}\text{ GeV}.
\end{equation}
With the operators in Eq. (\ref{eq3-1}) and the $\sigma(x)$ parametrization, the ALPs mass term is written as follows \cite{Carvajal:2015dxa}
\begin{equation}
m_{a}=|g|^{\frac{1}{2}}DM_{\textrm{Pl}}\lambda^{\frac{D}{2}-1},\label{eq11}
\end{equation}
where $10^{-10}\lesssim\lambda\equiv\frac{v_{\sigma}}{\sqrt{2}M_{\textrm{Pl}}}\lesssim10^{-5}$
and $M_{\textrm{Pl}}=2.44\times10^{18}$
GeV is the reduced Planck scale. 

Now, we turn our attention to the coupling of ALPs to photons which is determined by the interaction term $\frac{g_{a\gamma}}{4}a\,F_{\mu\nu}\widetilde{F}^{\mu\nu}$, where $F_{\mu\nu}$ and $\widetilde{F}^{\mu\nu}$ are the electromagnetic
field strength and its dual, respectively. This term is anomaly induced and given by \footnote{Higher corrections to the $g_{a\gamma}$ coupling are possible. For an extensive study of them to see \cite{Bauer:2017ris}. However, for the suitable ALPs masses in order to explain the observed DM relic density, all of them can be safely neglected.}
\begin{equation}
g_{a\gamma}=\frac{\alpha}{2\pi}\frac{C_{a\gamma}}{v_{\sigma}},
\end{equation}
with $\alpha\approx1/137$. Here, the electromagnetic
anomaly coefficient $C_{a\gamma}$ reads as \cite{Srednicki:1985xd,Marsh:2015xka}:
\begin{equation}
 C_{a\gamma}=2\sum_\psi\Big(X_{\psi_\textrm{L}}-X_{\psi_\textrm{R}}\Big)\Big(C_{\textrm{em}}^{(\psi)}\Big)^2\label{cagamma},
\end{equation}
where $C_{\textrm{em}}^{(\psi)}$ is the electric charge of the fermion $\psi$, and $X_{\psi_{L,R}}$ is its charge under the U$(1)_A$ symmetry. This anomaly coefficient is of order of one (1 or 2 more specifically) in our models and it directly determines the width of the red
band in Figure \ref{fig1} where ALPs are DM candidates.  Also, it is important to note that the existence of a non-null anomaly coefficient guarantees that $g_{a\gamma}\neq0$. This is the  reason for the total Lagrangian
in Eq. (\ref{eq1-1}) is invariant under an anomalous U$(1)_{A}$
global symmetry. Nevertheless, only with SM model fermions and the
neutral $S_{\textrm{R}\alpha}$ and $N_{\textrm{R}\beta}$ fermions
is not possible to have an anomalous U$(1)_{A}$ symmetry in the electromagnetic
group. Therefore, we need include the SU$(2)_{L}$ singlet fermion, $E$, with an unit of electric charge.

On the other hand,  the dimension $D$ of the gravity-induced mass operator in Eq. (\ref{eq3-1}) must be, in
general, larger than $4$ because the astrophysical and cosmological
constraints on the properties of ALPs. To be more specific, we show, in Figure \ref{fig1},
some regions of the ALPs space of parameters $- g_{a\gamma}\text{ vs }m_{a} -$
where ALPs give an explanation for some astrophysical anomalies and
others forbidden regions \cite{Graham:2015ouw,Budker:2013hfa,Graham:2013gfa,Ringwald:2012hr}.

Regarding the neutrino mass generation, we have that, once introduced
the $N_{\textrm{R}\beta}$ and $S_{\textrm{R}\alpha}$ fields, the
$\mathcal{L}_{\textrm{ISS}}$ Lagrangian reads as: 
\begin{widetext}
\begin{eqnarray}
\mathcal{L}_{\textrm{ISS}}=\ y_{i\beta}\overline{L_{i}}\widetilde{H}N_{\textrm{R}\beta}+\zeta_{\alpha\beta}\frac{\sigma^{p}}{M_{\textrm{Pl}}^{p-1}}\overline{S_{\textrm{R}\alpha}}(N_{\textrm{R}\beta})^{\textrm{C}}+\eta{}_{\alpha\alpha'}\frac{\sigma^{q}}{2M_{\textrm{Pl}}^{q-1}}\overline{S_{\textrm{R}\alpha}}(S_{\textrm{R}\alpha'})^{\textrm{C}} +\theta_{\beta\beta'}\frac{\sigma^{r}}{2M_{\textrm{Pl}}^{r-1}}\overline{N_{\textrm{R}\beta}}(N_{\textrm{R}\beta'})^{\textrm{C}}+\text{ H.c.},\label{eq7-1}
\end{eqnarray}
\end{widetext}
where the $y_{i\beta}$, $\zeta_{\alpha\beta}$, $\eta{}_{\,\alpha\alpha'}$,
$\theta_{\,\beta\beta'}$, coupling constants, with $i,j=1,2,3$,
$\alpha,\alpha'=1,2,(\text{or }3)$ and $\beta,\beta'=1,2,(\text{or }3)$,
are generically assumed of order one. The exponents $p,q,r$ are integer numbers chosen for satisfying some phenomenological constraints discussed
below. Negative values for these exponents will mean that the term is
$\sim\sigma^{*\,n}$ instead of $\sim\sigma^{n}$. Note that, without
loss of generality, the exponent $p$ can be assumed to be positive.
We will only consider the minimal number of neutral fermionic fields,
$S_{\textrm{R}\alpha}$ and $N_{\textrm{R}\beta}$, that allow to
fit all the experimental neutrino physics \cite{Abada:2014vea}. Specifically,
we study the $(2,2),$ $(2,3)$ and $(3,3)$ cases.

As the $\sigma$ field gets a VEV the
gravity-induced terms in Eq. (\ref{eq7-1}) give the mass matrix for
light (active) and heavy neutrinos \cite{Carvajal:2015dxa}. Specifically,
we can write the mass matrix in the $(\nu_{\textrm{L}},N_{\textrm{R}}^{\textrm{C}},S_{\textrm{R}}^{\textrm{C}})$
basis as 
\begin{align}
M_{\nu} & =\left[\begin{array}{cc}
0 & M_{D}^{\intercal}\\
M_{D} & M_{R}
\end{array}\right],\text{ with }\label{eq2}\\
M_{D}\equiv\left[\begin{array}{c}
m_{D}\\
0
\end{array}\right] & \text{ and }M_{R}\equiv\left[\begin{array}{cc}
\mu_{N} & M^{\intercal}\\
M & \mu_{S}
\end{array}\right].\label{eq3}
\end{align}
where $m_{D}$, $M$, $\mu_{N}$ and $\mu_{S}$ are matrices with
dimension equal to $n_{N_{R}}\times3$, $n_{N_{R}}\times n_{S_{R}}$,
$n_{N_{R}}\times n_{N_{R}},$ $n_{S_{R}}\times n_{S_{R}},$ respectively.
The energy scales of the entries in these matrices are determined
essentially by $\sqrt{2}\left\langle H\right\rangle \equiv v_{\text{SM}}\simeq246$
GeV, $\lambda$ (or $v_{\sigma}$) and $M_{\textrm{Pl}}$ GeV as follows
\begin{eqnarray}
m_{D\,i\beta}=y_{i\beta}\frac{v_{\text{SM}}}{\sqrt{2}}, & \  \ \  & M_{\alpha\beta}=\zeta_{\alpha\beta}M_{\textrm{Pl}}\lambda^{p},\label{eq4}\\
\mu_{S\,\alpha\alpha'}=\eta{}_{\alpha\alpha'}M_{\textrm{Pl}}\lambda^{\left|q\right|}, &  & \mu_{N\,\beta\beta'}=\theta_{\beta\beta'}M_{\textrm{Pl}}\lambda^{\left|r\right|}.\label{eq5}
\end{eqnarray}
The mass matrix in Eq. (\ref{eq2}) allows light active neutrino masses
at order of sub-eV without resorting very large energy scales in contrast
to the type I seesaw mechanism \cite{Minkowski:1977sc,GellMann:1980vs,Yanagida:1979as,Mohapatra:1979ia,Schechter:1980gr,Schechter:1981cv}. In more detail, assuming the hierarchy
$\mu_{N}\lesssim\mu_{S}\ll m_{D}<M$ (note that making $\mu_{S}$ and
$\mu_{N}$ small is technically natural) and taking a matrix expansion
in powers of $M^{-1}$, the light active neutrino masses, at leading
order, are approximately given by the eigenvalues of the matrix \cite{Grimus:2000vj,Boucenna:2014zba}
\begin{eqnarray}
m_{\nu\text{light}} & \simeq & m_{D}^{\intercal}M^{-1}\mu_{S}(M^{\intercal})^{-1}m_{D}.\label{eq6}
\end{eqnarray}
On the other hand, the heavy neutrino masses are given by the eigenvalues
of $m_{\nu\textrm{heavy}}\simeq M_{R}$. Note  from  Eq. (\ref{eq6}) that $\mu_{N}$ does
not contribute to the light active neutrino masses at the leading
order \cite{Grimus:2000vj,Boucenna:2014zba}. Actually, the presence
of $\mu_{N}$ term gives a subleading contribution to $m_{\nu\text{light}}$ of the order of $m_{D}^{\intercal}M^{-1}\mu_{S}(M^{\intercal})^{-1}\mu_{N}M^{-1}\mu_{S}(M^{\intercal})^{-1}m_{D}$, 
which is a factor $\mu_S\mu_N/M^2$ smaller than the leading contribution \cite{Carvajal:2015dxa}.

Very motivated scales for $M$ and $\mu_{S},\,\mu_{N}$ are TeV
and keV scales, respectively. These scales allow getting active neutrino masses in the sub-eV scale without considering  smaller Yukawas and, in some scenarios, such as the $(2,3)$ ISS case, the existence of a keV sterile neutrino as a warm dark matter
(WDM) candidate \cite{Abada:2014zra}. In addition, $M$ has to satisfy
\begin{equation}
M\gtrsim\sqrt{10\frac{\mu_{S}}{\textrm{keV}}}\ \textrm{TeV},
\end{equation}
because light active neutrino masses are in sub-eV scale and $m_{D}$
is of order of ${\cal O}(v_{\text{SM}})$. 

Another constraint on the $M$ scale comes from the fact that the mixing matrix that relates the three left-handed neutrinos with the three lightest mass-eigenstate neutrinos is not longer unitary. 
This implies that deviations of some SM observables may be expected, such as additional contributions to the $\ell \nu W$ vertex and to lepton-flavor and CP-violating processes, and non-standard effects in neutrino propagation \cite{Hettmansperger:2011bt, deSalas:2017kay}.
For example, in the inverse seesaw model, the violation of unitary is of order of  $\epsilon^{2}$, with $\epsilon\equiv m_{D}M^{-1}$ being approximately the mixing between light active and heavy neutrinos \cite{Boucenna:2014zba}.
Roughly speaking, $\epsilon^{2}$ at the percent level is not excluded experimentally \cite{Hettmansperger:2011bt,Ibarra:2010xw,PhysRevD.88.113001,Das:2014jxa}.

Taking into account the previous considerations, the ranges chosen
for $M$ and $\mu_{S}$ are 
\begin{equation}
1\leq M\leq25\ \textrm{TeV},\ 0.1\leq\mu_{S}\leq50\ \textrm{keV.}\label{eq8}
\end{equation}
Once established that scales of the mass matrices and using Eqs. (\ref{eq4}) and (\ref{eq5}) (and following a similar procedure as in Ref. \cite{Carvajal:2015dxa}), 
the integers $p$ and $q$ in Eq. (\ref{eq7-1}) can only take the
values 
\begin{eqnarray}
(p,|q|) & = & (2,3)\ \  \text{for}\  6\times10^{10}\lesssim v_{\sigma}\lesssim1\times10^{11}\ \ \textrm{GeV},\nonumber\\ \label{eq9}\\
(p,|q|) & = & (3,5)\ \ \text{for}\  2\times10^{13}\lesssim v_{\sigma}\lesssim8\times10^{13}\ \ \textrm{GeV}.\nonumber\\ \label{eq10}
\end{eqnarray}
That happens because the same VEV simultaneously provides $M$ and
$\mu_{S}$ scales. Note that for both possibilities in Eqs. (\ref{eq9}) and
(\ref{eq10}) the light active neutrino mass matrix in Eq. (\ref{eq6}) is simplified
to 
\begin{equation}
m_{\nu\textrm{light}}=\big[y^{\intercal}\zeta^{-1}\eta\left(\zeta^{\intercal}\right)^{-1}y\big]\frac{v_{\text{SM}}^{2}}{\sqrt{2}v_{\sigma}}.
\end{equation}
Moreover, the exponent $r$ of the term that generates $\mu_{N}$ in Eq. (\ref{eq7-1}) 
is also constrained to be $r\geq |q|$, because $\mu_{N}$ must be $\lesssim\mu_{S}$.

Finally, we have that $\mathcal{L}_{E}$, the Lagrangian involving
the $E$ charged fermion, is written as 
\begin{eqnarray}
\mathcal{L}_{E} & \supset & \vartheta_{i}\frac{\sigma^{s}}{M_{\textrm{Pl}}^{s}}\overline{L_{i}}HE_{\textrm{R}}+\kappa\frac{\sigma^{t}}{M_{\textrm{Pl}}^{t-1}}\overline{E_{\textrm{L}}}E{}_{\textrm{R}}+\textrm{H.c.},\label{eq18-1}
\end{eqnarray}
where $\vartheta_{i}$ and $\kappa$ are Yukawas, in principle, assumed of
order one. These two terms are also subjected to phenomenological
and theoretical constraints as follow. Because the term $\sim\sigma^{t}\overline{E_{\textrm{L}}}E{}_{\textrm{R}}$
must give a mass large enough for the $E$ fermion to satisfy its
experimental constrains. For stable charged heavy lepton, $m_{E}>102.6$ GeV at 95$\%$ C.L.  \cite{Olive:2016xmw},  or for charged long-lived heavy lepton, $m_{E}>574$ GeV at 95$\%$ C.L.  assuming mean life above $7\times 10^{-10}-3\times 10^{-8}$ s \cite{Chatrchyan:2013oca,doi:10.1142/S0217751X01003548}, $t$ must be less or equal than $3$. It must be different from zero because the electromagnetic anomaly must be present. 
On the other hand, $s$ can take the values $1$ or $2$ because $\sim\sigma^{s}\overline{L}HE_{\textrm{R}}$
determines the interaction of the $E$ fermion with the SM leptons, and whether $s$ is larger than $2$, the charged $E$ fermion becomes
stable enough to bring cosmological problems, unless its mass is $\lesssim\,$TeV.
Another constraint comes from searches for long-lived particles in
pp collisions \cite{Chatrchyan:2013oca,doi:10.1142/S0217751X01003548}.

Now an important discussion about the stability of both the ISS mechanism
and the ALPs mass is in order. In general, the gravitational effects
must be controlled to give a suitable ALPs mass. With this aim, we
introduced a gauge discrete $Z_{N}$ symmetry assumed as a remnant
of a gauge symmetry valid at very high energies \cite{PhysRevLett.62.1221}.
Thus, to truly protect the ALPs mass against those effects, $Z_{N}$ must
at least be free anomaly \cite{Ibanez1991291,PhysRevD.45.1424,Ibanez1993301,Luhn:2008sa},
i.e., 
\begin{equation}
A_{2}(Z_{N})=A_{3}(Z_{N})=A_{\textrm{grav}}(Z_{N})=0\hspace{0.5cm}\textrm{Mod}\ \frac{N}{2},
\end{equation}
where $A_{2},$ $A_{3}$ and $A_{\textrm{grav}}$ are the $[\text{SU}(2)_{\textrm{L}}]^{2}\times Z_{N}$,
$[\text{SU}(3)_{\textrm{C}}]^{2}\times Z_{N}$ and $[\textrm{gravitational}]^{2}\times Z_{N}$
anomalies, respectively. Other anomalies, such as $Z_{N}^{3}$, do not give useful low-energy constraints because these depend on some arbitrary choices concerning to the full theory.

Gravitational effects can also generate terms
such as $\frac{\sigma^{n}}{M_{\textrm{Pl}}^{n-1}}\overline{S_{\textrm{R}}}S_{\textrm{R}}^{\textrm{C}}$,
$\frac{\sigma^{n}}{M_{\textrm{Pl}}^{n-1}}\overline{S_{\textrm{R}}}N_{\textrm{R}}^{\textrm{C}}$,
$\frac{\sigma^{n}}{M_{\textrm{Pl}}^{n-1}}\overline{N_{\textrm{R}}^{\textrm{C}}}N_{\textrm{R}}$
or $\frac{\sigma^{n}}{M_{\textrm{Pl}}^{n}}\overline{L}\widetilde{H}S_{\textrm{R}}$
(with $n$ smaller than those in the Lagrangian (\ref{eq7-1})) that
jeopardize both the matrix structure - Eqs. (\ref{eq2}) and (\ref{eq3})
- and the scales of the ISS mechanism. Thus, $Z_{N}$ will be chosen
such that it also prevents these undesirable terms from appear.
\begin{table}
\setlength{\tabcolsep}{4pt} \global\long\def\arraystretch{1.5}
\begin{centering}
\begin{tabular}{c c c c c c c c c c c c}
\hline 
\hline
 & $Q_{\textrm{L}i}$  & $d_{\textrm{R}i}$  & $u_{\textrm{R}i}$  & $H_{i}$  & $L_{i}$  & $l_{\textrm{R}i}$  & $N_{\textrm{R}\beta}$  & $S_{\textrm{R}\alpha}$  & $E_{\textrm{L}}$  & $E_{\textrm{R}}$  & $\sigma$ \tabularnewline
\hline 
$\textrm{B}$  & $\frac{1}{3}$  & $\frac{1}{3}$  & $\frac{1}{3}$  & $0$  & $0$  & $0$  & $0$  & $0$  & $0$  & $0$  & $0$\tabularnewline

$\mathbb{L}$  & $0$  & $0$  & $0$  & $0$  & $1$  & $1$  & $1$  & $a$  & $b$  & $c$  & $d$\tabularnewline
\hline 
\hline
\end{tabular}
\par\end{centering}
\caption{Two of the continuous symmetries of the Lagrangian in Eq. (\ref{eq1-1}). $\textrm{B}$
and $\mathbb{L}$ are the baryon number and the generalized lepton
number, respectively. The charges $a$, $b$, $c$ and $d$ are given
by $a=qd/2,$ \textbf{$b=sd+c$}, $c=1-rd$ and $d=\left(p-q/2\right)^{-1}$.\label{table1} }
\end{table}

In general, the $Z_{N}$ symmetry can be written as a linear combination of the
continuous symmetries in the model: the hypercharge $Y$, the baryon
number $\textrm{B}$ and the generalized lepton number $\mathbb{L}$. The charge assignments for $\textrm{B}$ and $\mathbb{L}$ symmetries are shown in Table \ref{table1}, whereas the assignment for Y symmetry is the canonical one.
Nevertheless, since the hypercharge is free anomaly by construction,
the $Z_{N}$ charges $(Z)$ of the fields can be written as $Z=c_{1}\textrm{B}+c_{2}\mathbb{L}$,
where $c_{1,2}$ are rational numbers in order to make the $Z_{N}$
charges integers \cite{Carvajal:2015dxa}. Now, substituting the charges in Table \ref{table1} into the general form of the $Z_{N}$ symmetry (see Refs. \cite{Ibanez1991291,PhysRevD.45.1424,Ibanez1993301,Luhn:2008sa})
we can obtain the anomaly coefficients. Doing so, we find that $A_{3}(Z_{N})=0$
and 
\begin{eqnarray}
A_{2}(Z_{N}) & = & \frac{3}{2}\left[c_{1}+c_{2}\right],\label{eq13}\\
A_{\textrm{grav}}(Z_{N}) & = & c_{2}\Big[3-n_{N_{R}}-n_{S_{R}}\,\times\left[\frac{qd}{2}\right]+sd\Big].\label{eq14}
\end{eqnarray}
Note that $A_{2}(Z_{N})$ and $A_{\textrm{grav}}(Z_{N})$
are not, in general, $0$ Mod $N/2$ which implies strong constraints
on the choice of the $Z_{N}$ discrete symmetry.

\subsection{ALPs and sterile neutrino dark matter \label{sec22}}

Since the ALPs are very weakly interacting slim particles and cosmologically
stable, they can be considered as DM candidates \cite{Arias:2012az}. In fact, ALPs may
be nonthermally produced via the misalignment mechanism in the early
Universe and survive as a cold dark matter population until today.
Specifically, its relic density is determined from the following equation
\cite{PhysRevLett.103.111301,Sikivie:2006ni,2015arXiv150207375H,Bernabei:2003za,Arias:2012az,Masso:2006id,PhysRevD.82.123508,Marsh:2015xka}
\begin{equation}\label{ecu alp relic density}
\Omega_{a,\text{DM}}h^{2}\approx0.16\,\left[\frac{\Theta_{i}}{\pi}\right]^{2}\times\left[\frac{m_{a}}{\textrm{eV}}\right]^{1/2}\left[\frac{v_{\sigma}}{10^{11}\ \textrm{GeV}}\right]^{2},
\end{equation}
where $\Theta_{i}$ is the initial misalignment angle, which is taken
as $\frac{\pi}{\sqrt{3}}$, because we are assuming a post-inflationary
symmetry-breaking scenario, favorable for models with $v_{\sigma}\lesssim10^{14}$
GeV \cite{Arias:2012az,Sikivie:2006ni}.

On the other hand, the fraction of DM abundance in form of sterile
neutrino depends on its mass, $m_{\nu_{S}}$, and its mixing angle
with the light active neutrino, $\theta$. Specifically, $\nu_{S}$
as a WDM candidate can be generated through the well-known Dodelson-Widrow
(DW) mechanism \cite{Dodelson:1993je}, which is present as long as active-sterile mixing is
not zero \cite{Kusenko:2009up,Abazajian:2012ys,Adhikari:2016bei}. In the $(2,3)$ ISS
case, the sterile neutrino through the DW mechanism can account at maximum for $\approx 43\%$ of the
observed relic density without conflicting with observational constraints  \cite{Abada:2014zra}. This DM amount can be slightly increased to $\approx48\%$ when including
effect of the entropy injection of the pseudo-Dirac neutrinos provided
the lightest pseudo-Dirac neutrino has mass $1-10$ GeV \cite{Abada:2014zra}.
We are not going to consider these effects here. 
For $m_{\nu_{S}}>0.1$ keV, the relic density produced in the usual
DW mechanism is given by \cite{Asaka:2006nq,Abada:2014zra}
\begin{eqnarray}
\Omega_{\nu_{S},\textrm{DM}}h^{2}=&&\ 1.1\times10^{7}\sum_{\alpha}C_{\alpha}(m_{S})|U_{\alpha S}|^{2}\left[\frac{m_{\nu_{S}}}{\textrm{keV}}\right]^{2};\nonumber\\  &&\alpha=e,\mu,\tau,
\end{eqnarray}
where $C_{\alpha}(m_{S})$ are active flavor-dependent coefficients
which are calculated solving numerically the Boltzmann equations (an
appropriated value in this case is $C_{\alpha}(m_{S})\simeq0.8$ \cite{Asaka:2006nq}).
We also have that the sum of $U_{\alpha S}$, the elements of the
leptonic mixing matrix, is the active-sterile mixing, i.e., $\sum_{\alpha}|U_{\alpha S}|^{2}\sim\sin^{2}(2\theta)$.
For the case $m_{\nu_{S}}<0.1$ keV there is a simpler expression
written as follows \cite{Abazajian:2001nj,Abada:2014zra} 
\begin{equation}
\Omega_{\nu_{S},\textrm{DM}}h^{2}=0.3\left[\frac{\sin^{2}2\theta}{10^{-10}}\right]\left[\frac{m_{\nu_{S}}}{100\text{ keV}}\right]^{2}.
\end{equation}
After imposing bounds coming from stability,
structure formation and indirect detection, in addition to the constraints
arising from the neutrinos oscillation experiments, it was found that
the sterile neutrino as WDM in the $(2,3)$ ISS provides a sizable
contribution to the DM relic density for $2\lesssim m_{\nu_{S}}\lesssim50$
keV and active-sterile mixing angles $10^{-8}\lesssim\sin^{2}(2\theta)\lesssim10^{-11}$  \cite{Abada:2014zra}, where the maximal fraction of DM made of $\nu_{S}$ is achieved when $m_{\nu_{S}}\simeq7$
keV \cite{Boyarsky:2008ju,Boyarsky:2008xj,Horiuchi:2013noa,Boyarsky:2012rt}.

Once established the DM candidates and the parameters that determine the relic density in each case, we are going to search for models satisfying all mentioned conditions
in Section \ref{framework} as well as 
\begin{eqnarray}
\Omega_{\text{DM}}^{\text{Planck}}h^{2} & = & \Omega_{\nu_{S},\textrm{DM}}h^{2}+\Omega_{a,\text{DM}}h^{2},
\end{eqnarray}
where $\Omega_{\text{DM}}^{\text{Planck}}h^{2}=0.1197\pm0.0066$
(at $3\sigma$) is the current relic density as reported by Planck Collaboration
\cite{Ade:2015xua}.

\section{Models \label{model}}
\begin{figure*}
\center \includegraphics[scale=0.35]{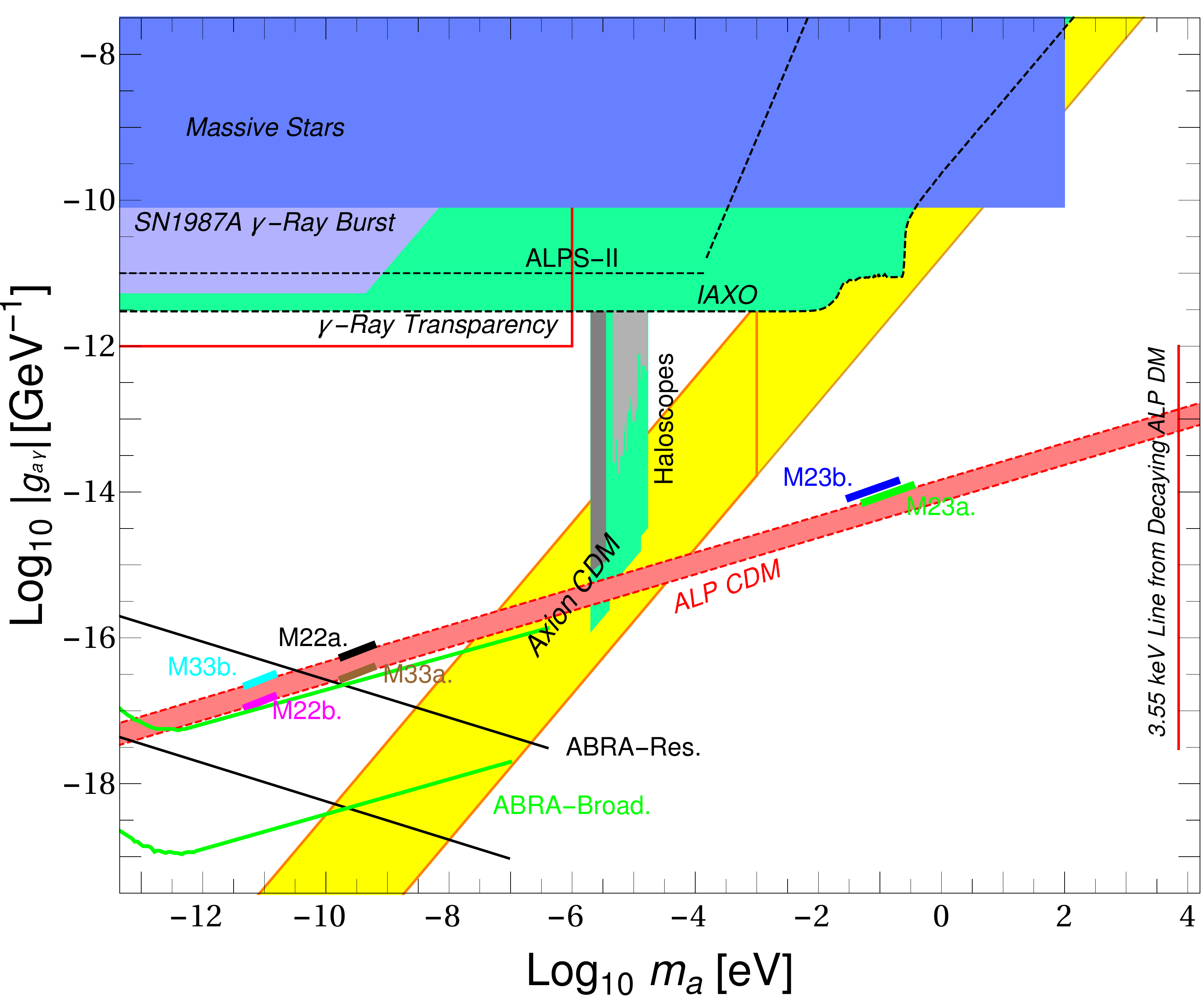}\caption{ALPs parameter space. This figure shows some excluded regions from
the nonobservation of an anomalous energy loss of massive stars due
to ALPs (or axions) emission \cite{Friedland:2012hj}, of a $\gamma$-ray
burst from SN 1987A due to conversion of an ALPs in the galactic magnetic
field \cite{Grifols:1996id,Brockway:1996yr,Payez:2014xsa} and of
dark matter axions or ALPs converted into photons in microwave cavities
placed in magnetic fields \cite{Bahre:2013ywa,PhysRevD.92.021101,Armengaud:2014gea,Rosenberg:2015kxa}.
It is also showed the red band where the ALPs may constitute all of cold dark matter (ALPs CDM), and the regions where the ALPs may explain
the cosmic $\gamma$-ray transparency and the x-ray line at 3.55 keV
\cite{Meyer:2013pny,Conlon:2013txa,Angus:2013sua,Higaki:2014zua,Jaeckel:2014qea}.
The green regions are the projected sensitivities of the light-shining-through-wall
experiment ALPS-II, of the helioscope IAXO, of the haloscopes ADMX
and ADMX-HF \cite{Bahre:2013ywa,Armengaud:2014gea}.
The black (green) solid line in the lower left corner shows the sensitivity of the proposed ABRACADABRA experiment \cite{Kahn:2016aff} using a resonant (broadband) circuit.
Furthermore, it is showed the region of QCD axion (the yellow band) -which was recently extended to cover the lower right corner \cite{PhysRevLett.118.031801}- studied in the context of some realistic axion models \cite{Olive:2016xmw}, with the region below the orange solid line corresponding to the axion CDM. The benchmark regions M$22$a, M$22$b, M$23$a, M$23$b, M$33$a and M$33$b corresponding to respective models $(2,2)$ ISS, $(2,3)$
ISS and $(3,3)$ ISS which generate a considerable amount of the current
DM relic density are also shown. \label{fig1}}
\end{figure*}
In the previous section, we have introduced the general and minimal constraints
that models have to satisfy. Now, we proceed to find specific
models that give an explanation to the dark matter observed in the
Universe. In particular, the $(2,2)$, $(3,3)$ and $(2,3)$ cases of the ISS mechanism	
are studied in detail.
For each model we check the compatibility (at 3$\sigma$) with the experimental neutrino physics \cite{deSalas:2017kay} for the normal mass ordering and vanishing CP phases by varing the free Yukawa couplings  $y_{i\beta}, \zeta_{\alpha\beta},\eta_{\alpha\alpha},\theta_{\beta\beta}$ in the range $\sim(0.1,3.5)$. Additionally, we also analize the lepton flavor violating processes such as $\ell_\beta\to \ell_\alpha+\gamma$, which are induced at one loop by the $W$ boson and the heavy neutrinos. 

The correspondig decay rates read \cite{GonzalezGarcia:1991be,Deppisch:2004fa} 
\begin{eqnarray}
  \mbox{Br}(\ell_\beta\to \ell_\alpha\gamma)=&&\ \frac{\alpha_{W}^3s_W^2}{256\pi^2}\frac{m_{\ell_\beta}^5}{m_W^4\Gamma_{\ell_\beta}}\nonumber\\ &&\times\left|\sum_iU_{\beta i}^*U_{\alpha i}G(m_{N_i}^2/m_W^2)\right|^2,
\end{eqnarray}

where $G(x)=x(1-6x+3x^2+2x^3-6x^2\log(x))/[4(1-x)^4]$, $\Gamma_{\ell_\beta}$ is the total decay width of $\ell_\beta$ and $U$ represents the lepton mixing matrix. 
We verify that each ISS model is compatible with the current experimental limits Br$(\mu\to e\gamma)<5.7\times10^{-13}$ \cite{Adam:2013mnn}, Br$(\tau\to e\gamma)<3.3\times10^{-8}$ and Br$(\tau\to \mu\gamma)<4.4\times10^{-8}$ \cite{Aubert:2009ag}. 

\subsection{$(2,2)$ ISS case\label{22case}}

Among the minimal configuration of the ISS mechanism consistent with
the experimental neutrino physics and lepton-flavor-violating (LFV) processes \cite{Abada:2014vea,Abada:2014zra,Das:2017ski} (for a recent review see Ref. \cite{Lindner:2016bgg}),
we, firstly, study the $(2,2)$ ISS case because this is the minimal configuration that satisfy all the constraints coming from experimental neutrino physics. For this case, in the neutrino mass spectrum 
there are two heavy pseudo-Dirac neutrinos with masses $\sim M$ and three light
active neutrinos with masses of order of sub-eV coming from the mass
matrix in Eq. (\ref{eq6}) \cite{Abada:2014vea}. Because in this case  $n_{N_{R}}=n_{S_{R}}=2$ (similarly for the $(3,3)$ ISS case) there is not a light sterile neutrino $\nu_{S}$ in the mass spectrum. Therefore, all the current DM abundance must be constituted
by ALPs, i.e. $\Omega_{\text{DM}}^{\text{Planck}}h^{2}=\Omega_{a,\text{DM}}h^{2}$. 

\begin{table}
\setlength{\tabcolsep}{10pt} \global\long\def\arraystretch{1.3}
\begin{centering}
\begin{tabular}{ c c c }
\hline 
\hline
$D$  & $v_{\sigma}$ $\left(\text{GeV}\right)$
& $v_{\sigma}$ $\left(\text{GeV}\right)$ 
\tabularnewline
\hline 
8  & $(1.9-3.3)\times10^{10}$  & $(1.5-2.8)\times10^{10}$ \tabularnewline

9  & $(0.6-1.1)\times10^{11}$  & $(5.5-9.5)\times10^{10}$\tabularnewline

10  & $(1.9-3.2)\times10^{11}$  & $(1.6-2.8)\times10^{11}$ \tabularnewline

11  & $(5.0-8.2)\times10^{11}$  & $(4.4-7.2)\times10^{11}$ \tabularnewline

12  & $(1.2-1.9)\times10^{12}$  & $(1.0-1.7)\times10^{12}$\tabularnewline

13  & $(2.6-4.0)\times10^{12}$  & $(2.3-3.6)\times10^{12}$\tabularnewline

14  & $(5.2-7.9)\times10^{12}$  & $(4.6-7.0)\times10^{12}$ \tabularnewline

15  & $(1.0-1.4)\times10^{13}$  & $(0.9-1.3)\times10^{13}$\tabularnewline

16  & $(1.7-2.5)\times10^{13}$  & $(1.6-2.3)\times10^{13}$ \tabularnewline

17  & $(2.9-4.2)\times10^{13}$  & $(2.7-3.8)\times10^{13}$ \tabularnewline

18  & $(4.8-6.7)\times10^{13}$  & $(4.3-6.1)\times10^{13}$ \tabularnewline

19  & $(0.7-1.0)\times10^{14}$  & $(6.8-9.4)\times10^{13}$ \tabularnewline
\hline 
\hline
\end{tabular}
\par\end{centering}
\protect\caption{The appropriated $\left(D,\,v_{\sigma}\right)$ values in order to
provide $\Omega_{a,\text{DM}}h^{2}=\Omega_{\text{DM}}^{\text{Planck}}h^{2}$
(first and second columns) and $\Omega_{a,\text{DM}}h^{2}=0.57\times\Omega_{\text{DM}}^{\text{Planck}}h^{2}$
(first and third columns). It has been considered
$10^{-3}\leq g\leq2$ and $\Omega_{\textrm{DM}}^{\textrm{Planck}}h^{2}$
at 3$\sigma$ level.\label{table2}}
\end{table}
In order to find the main features of the model, we find useful to
rewrite $\Omega_{a,\text{DM}}h^{2}$ in terms of $D$ - the exponent
 of the mass operator for $\sigma$, Eq. (\ref{eq3-1}) - and $m_{a}$. Thus, substituting Eqs.(\ref{eq3-1}) and  (\ref{eq11}) in Eq. (\ref{ecu alp relic density}), we find that 
\begin{eqnarray}
\Omega_{a,\text{DM}}h^{2}\simeq&&\ 0.49\,|g|^{\frac{1}{4}}\,\sqrt{D}\exp\left[-\frac{D}{4}\ln\frac{\sqrt{2}M_{\text{Pl}}}{1\text{ GeV}}\right]\nonumber\\ &&\times\left[\frac{v_{\sigma}}{1\text{ GeV}}\right]^{\frac{D+6}{4}},
\end{eqnarray}
where $g$ is assumed to be $10^{-3}\leq\left|g\right|\leq2$.
Thus, we can see that $\Omega_{a,\text{DM}}h^{2}$ only depends on $\left(\left|g\right|,\,D,\,v_{\sigma}\right)$.
In Table \ref{table2} we show $\left(D,\,v_{\sigma}\right)$ values
for the cases where $\Omega_{a,\text{DM}}h^{2}=\Omega_{\text{DM}}^{\text{Planck}}h^{2}$
and $\Omega_{a,\text{DM}}h^{2}=0.57\times\Omega_{\text{DM}}^{\text{Planck}}h^{2}$.
The last case applies only for the $(2,3)$ ISS case and will be discussed
in Section \ref{23case}.

In order to obtain the Lagrangian in this scenario,
we search for discrete symmetries for the two possibilities showed in
Eqs. (\ref{eq9}-\ref{eq10})  and different values of $r,s,t$ according its respective constraints as follow: considering Eqs. (\ref{eq9}-\ref{eq10}) and the Table \ref{table2} we can see that, for the range of values of $v_\sigma$ established in the Section \ref{sec21}, only the values  $D=9,10,16-19$ are allowed to reproduce the correct relic density to ALPs. Thus we searched for discrete symmetries $Z_N$ that allows the mass operators with those dimensions $D$,  with the following results:
The $Z_{9,10,}$ symmetries allow terms such as $\frac{\sigma^{*}}{M_{\textrm{Pl}}}\bar{L}\widetilde{H}S_{R}$,
$\frac{\sigma^{*2}}{M_{\textrm{Pl}}}\bar{N}_{R}N_{R}^{C}$, $\bar{L}\widetilde{H}S_{R}$, and since $H$ and $\sigma$ get VEVs, these terms do
not give the appropriate zero texture of the ISS mechanism shown in
Eqs. (\ref{eq2}) and (\ref{eq3}).  We have searched for all the possible
combinations of $r,s,t$ values in the Lagrangian of Eq. (\ref{eq7-1}) without any success. On the other hand, the $Z_{16,18}$ symmetries
are not free of the gravitational anomaly. In fact, the $Z_{N\leq20}$
discrete symmetries that satisfy all the anomaly constraints and stabilize
the ISS mechanism are $Z_{17,19}$. 

In the case of $Z_{17}$ symmetry, the Lagrangian, $\mathcal{L}{}_{Z_{17}}$, is given
by Eq. (\ref{eq1-1}) with the parameters $D=17$ and $(p,q,r,s,t)=(3,-5,-6,2,2)$
in Eqs. (\ref{eq3-1}), (\ref{eq7-1}) and (\ref{eq18-1}), respectively. An assignment of the $Z_{17}$ (with $Z_{17}=6\text{B}+11\text{U}(1)_{\mathbb{L}}$)
charges and the anomalous U$(1)_{A}$ symmetry for this case is
shown in Table \ref{table9-1}. Note that, for this model the
term $\sim\sigma^{*6}\overline{N_{\textrm{R}\beta}}(N_{\textrm{R}\beta'})^{\textrm{C}}$ in Eq. (\ref{eq7-1})
gives a negligible contribution for the light active neutrino masses.

\begin{table*}
\setlength{\tabcolsep}{0.15cm} \global\long\def\arraystretch{1.5}
\begin{centering}
\begin{tabular}{c c c c c c c c c c c c c}
\hline 
\hline
 Model & Symmetry & $Q_{\textrm{L}i}$  & $d_{\textrm{R}i}$  & $u_{\textrm{R}i}$  & $H$  & $L_{i}$  & $l_{\textrm{R}i}$  & $N_{\textrm{R}\beta}$  & $S_{\textrm{R}\alpha}$  & $E_{\textrm{L}}$  & $E_{\textrm{R}}$  & $\sigma$ \tabularnewline
\hline
 \multirow{4}{0.8cm}{(2,2)} & $Z_{17}$  & $1$  & $2$  & $0$  & $16$  & $14$  & $15$  & $13$  & $8$  & $15$  & $12$  & $7$\\
&U$(1)_{A}$  & $0$  & $0$  & $0$  & $0$  & $11/2$  & $11/2$  & $11/2$  & $-5/2$  & $11/2$  & $15/2$  & $1$\\
&$Z_{19}$  & $1$  & $14$  & $7$  & $6$  & $16$  & $10$  & $3$  & $9$  & $17$  & $2$  & $4$\\
&U$(1)_{A}$  & $0$  & $0$  & $0$  & $0$  & $11/2$  & $11/2$  & $11/2$  & $-5/2$  & $5/2$  & $7/2$  & $1$\\\cline{1-13}

\multirow{4}{0.8cm}{(3,3)} & $Z_{17}$  & $1$  & $10$  & $9$  & $8$  & $14$  & $6$  & $5$  & $7$  & $2$  & $15$  & $4$\\
&U$(1)_{A}$  & $0$  & $0$  & $0$  & $0$  & $11/2$  & $11/2$  & $11/2$  & $-5/2$  & $9/2$  & $7/2$  & $1$\\
&$Z_{19}$  & $1$  & $13$  & $8$  & $7$  & $16$  & $9$  & $4$  & $12$  & $9$  & $11$  & $18$\\
&U$(1)_{A}$  & $0$  & $0$  & $0$  & $0$  & $11/2$  & $11/2$  & $11/2$  & $-5/2$  & $11/2$  & $7/2$  & $1$\\\cline{1-13}

\multirow{3}{0.8cm}{(2,3)} & $Z_{10}$  & $1$  & $2$  & $0$  & $9$  & $7$  & $8$  & $6$  & $6$  & $8$  & $6$  & $6$\\
&$Z_{4}$  & $0$  & $1$  & $3$  & $3$  & $0$  & $1$  & $3$  & $1$  & $1$  & $1$  & $2$\\
&U$(1)_{A}$  & $0$  & $0$  & $0$  & $0$  & $7/2$  & $7/2$  & $7/2$  & $-3/2$  & $7/2$  & $3/2$  & $1$\tabularnewline
\hline 
\hline
\end{tabular}
\par\end{centering}
\protect\caption{Discrete and continuous charge assignments of the fields in the different
models.\label{table9-1}}
\end{table*}

The  corresponding $g_{a\gamma}$ and $m_{a}$ for this model is given  by 
\begin{eqnarray}
&&g_{a\gamma}  \cong\ 7.54\times10^{-17}\left[\frac{3.08\times10^{13}\textrm{ GeV}}{v_{\sigma}}\right]\textrm{ GeV}^{-1},\nonumber \\
&&m_{a}  \cong  5.59\times10^{-10}|g|^{\frac{1}{2}}\left[\frac{v_{\sigma}}{3.08\times10^{13}\textrm{ GeV}}\right]^{15/2}\,\textrm{eV}.\nonumber\\
\end{eqnarray}
The benchmark region for this case is denoted as M$22$a in the Figure \ref{fig1}
where we have considered $10^{-3}\leq\left|g\right|\leq2$
and $2.9\times10^{13}\lesssim v_{\sigma}\lesssim4.2\times10^{13}$
GeV. These values for $g_{a\gamma}$ and $m_{a}$
allow that the ALPs explain $100\%$ of the DM relic density. 

Sharp predictions for neutrinos masses are not possible with just
the knowledge of the $p,q,r,s,t$ values and $v_{\sigma}$. However,
the order of magnitude of the mass matrices can be estimated from
Eqs. (\ref{eq4}) and (\ref{eq5}) to be (using $v_\sigma\cong3.08\times10^{13}\textrm{ GeV}$)
\begin{eqnarray}
&&M\cong\zeta\times1.73\ \textrm{TeV}, \ \ \ \mu_{S}\cong\eta\times0.13\ \textrm{keV},\nonumber \\  &&\ m_{\nu\textrm{light}}\simeq\big[y^{\intercal}\zeta^{-1}\eta\left(\zeta^{\intercal}\right)^{-1}y\big]\times1.38\ \textrm{eV},\label{eq30}
\end{eqnarray}
which is appropriate to satisfy the constraints coming from experimental neutrino physics and unitarity without resorting a fine tuning in
couplings. Nevertheless, we have to admit that some care must be taken
in order to generate the benchmark region M$22$a in agreement with
bounds coming from LFV processes such
as $\mu\rightarrow e+\gamma$. 
Specifically, due to $m_{N_i}\sim M\gg m_W$ the loop function tends to $G(x)\to 1/2$ and the mixing terms are generically given by $U\sim m_D/M$. This leads to the decay rate for $\mu\rightarrow e+\gamma$ of the order of  
\begin{align}\label{eq:brmueg2}
\mbox{Br}(\mu\to e\gamma)\sim 1.1\times10^{-13}\left(\frac{m_D}{10\,\mbox{GeV}}\right)^4\left(\frac{3\,\mbox{TeV}}{M}\right)^4,
\end{align}
which implies that small $y\sim0.1$ couplings must be required.

For the case with $Z_{19}$, the effective Lagrangian is characterized by $(p,q,r,s,t)=(3,-5,-8,-2,1)$, and the results, roughly
speaking, are quite similar to the model with $Z_{17}$, in the sense that as the $p,q,|s|$ values are equals for both models, the neutrino spectrum is similar in both cases. Nevertheless, since $D,t$ values are not equals, we have as a consequence that: the ALPs mass, the mass term for the exotic fermion $E$ and the ALPs-photon coupling, $g_{a\gamma}$, are different. Specifically, from the Table \ref{table9-1} and Eq. (\ref{cagamma}), in the $Z_{19}$ model the ALPs parameters are 
\begin{eqnarray}
&&g_{a\gamma}  \cong 1.12\times10^{-17}\left[\frac{1.0\times10^{14}\textrm{ GeV}}{v_{\sigma}}\right]\textrm{ GeV}^{-1}, \nonumber \\ 
&&m_{a}  \cong  1.87\times10^{-10}|g|^{\frac{1}{2}}\left[\frac{v_{\sigma}}{1.04\times10^{14}\textrm{ GeV}}\right]^{17/2}\,\textrm{eV}.\nonumber\\
\end{eqnarray}

The benchmark region in this model corresponding to this case in Figure \ref{fig1} is denoted as M$22$b. We also show the values for the neutrino mass spectrum in Table \ref{table8}. 
Concerning to the upper bound on $\mu\rightarrow e+\gamma$, it is easily fulfilled due to the larger supression coming from $M\sim 50$ TeV. 

\subsection{$(3,3)$ ISS case\label{33case}}

Regarding the neutrino mass spectrum  the $(3,3)$ ISS case is quite
similar to the previous one in the sense that there is not a light sterile neutrino in the mass spectrum  because $n_{N_{R}}=n_{S_{R}}=3$.
Therefore, all the DM abundance in this model has to be made of ALPs.

Proceeding in a similar manner to the (2,2) ISS  case and taking into
account that $A_{\textrm{grav}}(Z_{N})$ is now different (see Eq. (\ref{eq14})), we have
searched for all anomaly-free $Z_{N}$ discrete symmetries, with $N\leq20$ and with $(p,\,q,\,r,\,s,\,t)$ values established according to the constraints in Section \ref{sec21}. Doing that, we found the following results:
the $Z_{9}$ symmetry is not free
of gravitational anomalies, while the $Z_{10}$ symmetry allows dangerous
terms such as $\bar{L}\widetilde{H}S_{R}$, $\frac{\sigma^{*}}{M_{\textrm{Pl}}}\bar{L}\widetilde{H}S_{R}$,
$\sigma\bar{N}_{R}N_{R}^{C}$, and others that
jeopardize the matrix structure in Eqs. (\ref{eq2}) and (\ref{eq3}), therefore the possibility
of building a model for the solution in Eq. (\ref{eq9}) is
not realized. On the other hand, the $Z_{16,18}$ symmetries corresponding
to the solution in Eq. (\ref{eq10}), are not free of gravitational
anomalies, therefore these are not suitable symmetries. However, the $Z_{17,19}$
symmetries forbid the dangerous terms and allow an effective
Lagrangian. 

In the case of $Z_{17}$ symmetry, the Lagrangian in Eq. (\ref{eq1-1}) is characterized
by the parameters $(p,\,q,\,r,\,s,\,t)=(3,\,-5,\,7,\,2,\,1)$ and $D=17$. Note then that this model has 
a Lagrangian very similar to
the $(2,2)$ ISS Lagrangian. However, in this case, the mass term for
the exotic fermion $E$ has the exponent equal to one and the term
associated with $\mu_{N}$ is not allowed with dimension less than
seven. Because the parameters $(p,\,|q|)$ are equals in both cases, the neutrino spectrum is the same as in the M22a model (see Eqs. (\ref{eq30})). Moreover, note that in this case, the
term $\sim\sigma^{7}\overline{N_{\textrm{R}\beta}}(N_{\textrm{R}\beta'})^{\textrm{C}}$
gives a negligible contribution for the light active neutrino masses. On the other hand,
the fact that the mass term for
the exotic fermion differs from (2,2) ISS model imply that the anomaly coefficient $C_{a\gamma}$ be different (see charges in the Table \ref{table9-1} and Eq. (\ref{cagamma})), such that the ALPs-photon coupling has also a different value.  Possible assignments for the $Z_{17}$ and U$(1)_{A}$ symmetries
are shown in Table \ref{table9-1},
with $Z_{17}=9\text{B}+11\text{U}(1)_{\mathbb{L}}$.

The corresponding $v_{\sigma}$ value
is the same that in the $(2,2)$ ISS case showed in the Table \ref{table2} for $D=17$, implying also that the $m_a$ is equal to it given in Eq. (\ref{eq30}). Nevertheless,
the $g_{a\gamma}$ turn to be 
\begin{eqnarray}
g_{a\gamma} & \cong & 3.77\times10^{-17}\left[\frac{3.08\times10^{13}\textrm{ GeV}}{v_{\sigma}}\right]\textrm{ GeV}^{-1},
\end{eqnarray}
because the anomaly coefficient now has a different value. A benchmark region for this case is denoted as M$33$a in Figure \ref{fig1}, where these values for $g_{a\gamma}$ and $m_{a}$ allow that the ALP explains
$100\%$ of the DM relic density. 

For the $Z_{19}$ case, we find that the model is determined by the parameters $(p,\,q,\,r,\,s,\,t)=(3,\,-5,\,-8,\,2,\,2)$, which brings
similar conclusions that the M22b model, with some differences coming from the anomaly coefficient $C_{a\gamma}$. Specifically, the coupling 
\begin{eqnarray}
g_{a\gamma}  \cong 2.24\times10^{-17}\left[\frac{1.0\times10^{14}\textrm{ GeV}}{v_{\sigma}}\right]\textrm{ GeV}^{-1}.
\end{eqnarray} 
The other parameters associated to neutrino spectrum and $m_a$ are similar than in the M22b case, and are shown in Table \ref{table8}.  
The benchmark region for this case is denoted as M$33$b in Figure \ref{fig1}. 

On the other hand, the constraints and prospects regarding lepton flavor violating processes are similiar to the ones in case (2,2) since the mass scale $M$ of the benchmark regions M33a and M33b are the same of the benchmark regions M22a and M22b, respectively.

We remark that a similar effective Lagrangian for the $(3,3)$
ISS case was worked in the Ref. \cite{Carvajal:2015dxa} with the
aim of explaining some astrophysical phenomena. However, in that case,
the DM abundance via ALP was not considered.

\subsection{$(2,3)$ ISS case\label{23case}}
For this case, because there are $n_{N_{R}}=2$ and $n_{S_{R}}=3$ neutral fermions,
the neutrino mass spectrum contains two 
heavy pseudo-Dirac neutrinos with masses $\sim M$ and three light 
active neutrinos with masses of order of sub-eV.  In addition, there is a sterile neutrino, $\nu_{S}$, with mass
of order $\sim\mu_{S}$. Then, for this model, the presence of
both the $\nu_{S}$ and the ALPs, $a$, brings the possibility of having
two DM candidates in the $(2,3)$ scenario \cite{Abada:2014zra,Dev:2012bd}. 

First, let's consider the case $\Omega_{\text{DM}}^{\text{Planck}}h^{2}=\Omega_{a,\text{DM}}h^{2}$,
i.e., when the DM abundance is totally constituted by ALPs. Now, from
Eqs. (\ref{eq9}) and (\ref{eq10}) and Table \ref{table2}, we can see
that  $\left(D,\,v_{\sigma}\right)=\left(9,\,(0.6-1.1)\times10^{11}\text{ GeV}\right)$
and $\left(D,\,v_{\sigma}\right)=\left(10,\,(1.9-3.2)\times10^{11}\text{ GeV}\right)$
corresponds to the $(p,|q|)=(2,3)$ solution in Eq. (\ref{eq9}) (note
that $v_{\sigma}$ corresponding to $D=10$ is slightly out of allowed
range in Eq. (\ref{eq9})). Moreover, the values $D=9,10$ restrict
the symmetry to be $Z_{9,10}$.
For these discrete symmetries we find solutions for anomaly free $Z_{9\text{ and }10}$
charges, i.e, solutions to Eqs. (\ref{eq13}) and (\ref{eq14}) with $(p,|q|)=(2,3)$.
Nevertheless, all the solutions for the $Z_{9\text{ and }10}$ charges
allow terms such as $\sim\sigma\overline{N_{\textrm{R}\beta}}(N_{\textrm{R}\beta'})^{\textrm{C}}$, $\sim\frac{\sigma^{*2}}{M_{\textrm{Pl}}}\overline{N_{\textrm{R}\beta}}(N_{\textrm{R}\beta'})^{\textrm{C}}$, $\sim\frac{\sigma^{*}}{M_{\textrm{Pl}}}\overline{L_{i}}\widetilde{H}S_{\textrm{R}\alpha}$ and other terms in the Lagrangian that do not give the correct texture to the mass matrix
in the ISS mechanism. We also have searched for all the possible
combinations of $r,s,t$ values in the Lagrangian (\ref{eq1-1}) with
$(p,|q|)=(2,3)$ without any success. Therefore, the $(p,|q|)=(2,3)$
case cannot offer a realization for an effective model providing
all the observed DM abundance via ALPs when all the constraints in
Section  \ref{framework} are considered. However, from Table \ref{table2}
we see that for  $D=15,\dots,19$ with a larger value
of $v_{\sigma}$ the second solution $(p,|q|)=(3,5)$, cf. Eq. \ref{eq10},
can, in principle, offer a model (note that, strictly speaking, the
$v_{\sigma}$ value corresponding to $D=15$ is slightly out of allowed
range in Eq. (\ref{eq10})). Moreover, the cases
of $Z_{17}$ and $Z_{19}$ are excluded because the condition for
the gravitational anomaly is never satisfied, while in the $Z_{16,18}$
cases terms as $\sim\overline{L_{i}}\widetilde{H}S_{\textrm{R}\alpha}$
and $\sim\sigma\overline{N_{\textrm{R}\beta}}(N_{\textrm{R}\beta'})^{\textrm{C}}$
give an incorrect texture for the ISS mass matrix. In fact, after
imposing all the constraints, we find that the only symmetry that
provides a solution is $Z_{15}$. In more detail, we find that the
discrete symmetry can be written as $Z_{15}=9\text{B}+11\text{U}(1)_{\mathbb{L}}$
(other combinations for $Z_{15}$ are possible). This model has the effective Lagrangian, $\mathcal{L}{}_{Z_{15}}$,
given by Eqs. (\ref{eq1-1}), (\ref{eq7-1}) and (\ref{eq18-1}) with
$(p,q,r,s,t)=(3,-5,-4,\,2,\,2)$. Note that the
term $\sim\sigma^{*4}\overline{N_{\textrm{R}\beta}}(N_{\textrm{R}\beta'})^{\textrm{C}}$
gives a negligible contribution for the light active neutrino masses.
  
The $\mathcal{L}{}_{Z_{15}}$ is also invariant under a U$(1)_{A}$
symmetry which is anomalous in the electromagnetic group as must be
to generate a non-null coupling between photons and ALPs, $g_{a\gamma}$ (see Sec. \ref{sec21}). Specifically,
for this case 
we have that the ALPs parameters are given by
\begin{eqnarray}
&&g_{a\gamma} \simeq 2.25\times10^{-16}\left[\frac{1.03\times10^{13}\textrm{ GeV}}{v_{\sigma}}\right]\textrm{ GeV}^{-1},\nonumber \\
&&m_{a} \simeq  4.47\times10^{-8}|g|^{\frac{1}{2}}\left[\frac{v_{\sigma}}{1.03\times10^{13}\textrm{ GeV}}\right]^{13/2}\,\textrm{eV.}\nonumber\\
\end{eqnarray}
We also check that the neutrino mass spectrum for this model is
\begin{eqnarray}
&&M\simeq\zeta\times6.5\times10^{-2}\ \textrm{TeV};\, \ \ \mu_{S}\simeq\eta\times5.8\times10^{-4}\ \textrm{keV};\nonumber \\ 
&&\ \ \ \ \ \ \ \ \ m_{\nu\textrm{light}}\simeq\big[y^{\intercal}\zeta^{-1}\eta\left(\zeta^{\intercal}\right)^{-1}y\big]\times4.15\ \textrm{eV},\label{eq23}
\end{eqnarray}
where we have used the particular value $v_{\sigma}\simeq1.03\times10^{13}$ GeV, which
is one of the suitable values given in Table \ref{table2} for $D=15$ giving the $100\%$ of the current DM abundance.  For this case, the sterile neutrino as DM candidate has a negligible contribution because the small scales in Eq. (\ref{eq23}) imply that the mixing angle between  the active and sterile neutrinos has a great suppression. Moreover the mass scale of the sterile neutrino, $\mu_S$, is very small to bring a considerable contribution to DM.

Now, from values of $M,\,\mu_{S},\,m_{\nu\textrm{light}}$ in Eq. (\ref{eq23}) we
note that in this scenario there is a some tension to satisfy the
unitarity constraint. In more detail, $\left|\frac{y}{\zeta}\right|<\frac{M}{v_{\text{SM}}}\times10^{-1}=2.6\times10^{-2}$
where we have been conservative choosing a $\epsilon^2$ value of $1\%$
(recall $\epsilon\equiv m_{D}M^{-1}$). However,
this upper bound on $\left|\frac{y}{\zeta}\right|$ implies a lower
bound on $\eta>\left|\frac{y}{\zeta}\right|^{-2}\frac{m_{\nu\textrm{light}}}{4.15}\approx\left|\frac{y}{\zeta}\right|^{-2}\frac{\sqrt{\Delta m_{\text{atm}}^{2}}}{4.15}\approx17.17$
(with $\Delta m_{\text{atm}}^{2}=2.32\times10^{-3}$ eV$^{2}$ being
the atmospheric squared-mass difference) which is not a perturbative
value for $\eta$. This happens because the values for $v_{\sigma}$
corresponding for $D=15$ is smaller than the values allowed in the
range in Eq. (\ref{eq10}). Similar conclusions are found if we consider
the case when $\Omega_{a,\text{DM}}h^{2}<\Omega_{\text{DM}}^{\text{Planck}}h^{2}$.
Therefore, the effective Lagrangian $\mathcal{L}{}_{Z_{15}}$ can
not provide a natural framework for DM and the neutrino masses in
$(2,3)$ ISS case. For this reason we do not show the benchmark
region for this model in Figure \textcolor{black}{\ref{fig1}.}

However, models explaining the DM relic density via ALPs and/or sterile
neutrinos for the $\left(2,3\right)$ ISS case can be found provided
we slightly relax some constraints mentioned in Section \ref{framework}.
Actually, if an extra $Z_{N}$ symmetry is allowed, we found that,
for example, the solution $\left(p,\left|q\right|\right)=\left(2,3\right)$ in Eq. (\ref{eq9})
makes possible a model with $D=10$ and $(p,q,r,s,t)=(2,-3,-3,2,2)$
in Eqs. (\ref{eq1-1}), (\ref{eq7-1}) and (\ref{eq18-1}),
where the discrete gauge symmetry $Z_{10}\times Z_{4}$, with the corresponding
charges given in Table \ref{table9-1}, must be considered with the aim of get the correct DM relic density using $D=10$ to calculate the ALPs mass.
It is straightforward to check that for this model, ALPs provide $100\%$
of the DM abundance provided\textcolor{red}{{} }\textcolor{black}{$v_{\sigma}\cong2.03\times10^{11}$}
GeV with $g$ of order one. In more details, for this benchmark point,
we have that 
\begin{eqnarray}
&&g_{a\gamma} \simeq 1.14\times10^{-14}\left[\frac{2.03\times10^{11}\textrm{ GeV}}{v_{\sigma}}\right]\textrm{ GeV}^{-1},\nonumber \\
&&m_{a} \simeq  0.29|g|^{\frac{1}{2}}\left[\frac{v_{\sigma}}{2.03\times10^{11}\textrm{ GeV}}\right]^{4}\,\textrm{eV.}
\end{eqnarray}
with the neutrino mass spectrum given by 
\begin{eqnarray}
&&M\simeq\zeta\times8.4\ \textrm{TeV};\,\ \ \ \ \ \mu_{S}\simeq \Big[\frac{\eta}{10^{-2}}\Big]\times4.96\ \textrm{keV};\nonumber\\
&&\ \ \,m_{\nu\textrm{light}}\simeq\big[y^{\intercal}\zeta^{-1}\Big(\frac{\eta}{10^{-2}}\Big)\left(\zeta^{\intercal}\right)^{-1}y\big]\times 2.11\ \textrm{eV}.
\end{eqnarray}
We note that for $\eta\leq10^{-2}$ and the other coupling constants of order one, a suitable neutrino mass spectrum is achieved. 

In this case, we have also check that the sterile neutrino has a negligible contribution to DM relic density because the mixing angle between the active and sterile neutrinos is smaller than the limits established to consider $\nu_S$ as a DM candidate ($10^{-8}\lesssim\sin^2(2\theta)\lesssim10^{-11}$, see ref. \cite{Abada:2014zra} for more details). For this model, we have shown in Figure \ref{fig1} a benchmark region denoted as M$23$a where ALPs provide $100\%$ of DM abundance. 

For the case that the DM abundance is made of ALPs and sterile neutrinos, the scenario slightly changes. We have chosen the case when the DM is made of $\approx43\%$ of sterile neutrinos and $\approx57\%$ of ALPs as an illustrating example. However, these can take other values provided the DM abundance made of sterile neutrinos is  $\lessapprox50\%$, consistently with the constraints over its parameter space \cite{Abada:2014zra,Merle:2015vzu,Boyarsky:2008xj}. Doing a similar procedure as in the previous cases, we can obtain
\begin{eqnarray}
&&g_{a\gamma} \simeq 1.02\times10^{-14}\left[\frac{2.28\times10^{11}\textrm{ GeV}}{v_{\sigma}}\right]\textrm{ GeV}^{-1},\nonumber\\
&&m_{a} \simeq  0.46|g|^{\frac{1}{2}}\left[\frac{v_{\sigma}}{2.28\times10^{11}\textrm{ GeV}}\right]^{4}\,\textrm{ eV.}
\end{eqnarray}
and 
\begin{eqnarray}
&&M\simeq\zeta\times10.6\ \textrm{TeV};\,\ \ \ \mu_{S}\simeq\Big[\frac{\eta}{10^{-2}}\Big]\times7.1\ \textrm{keV};\nonumber \\
&&\ \ m_{\nu\textrm{light}}\simeq\big[y^{\intercal}\zeta^{-1}\Big(\frac{\eta}{10^{-2}}\Big)\left(\zeta^{\intercal}\right)^{-1}y\big]\times1.9\ \textrm{eV}.
\end{eqnarray}
In this case for $\eta\approx10^{-2}$ the sterile neutrino has $m_{\nu_{S}}\approx7.1\,$KeV. In particular, this mass for the sterile neutrino may explain the recently indicated emission lines at 3.5 keV from galaxy clusters and the Andromeda galaxy \cite{Bulbul:2014sua,Boyarsky:2014jta}. The benchmark region for
this model is denoted as M$23$b in Figure \ref{fig1}. 

It is worth to mention that for both benchmark regions in those models, the constraints and prospects regarding lepton-flavor-violating processes are also similiar to the ones in case (2,2). This happens because the contribution of the sterile neutrino to Br$(\ell_\beta\to\ell_\beta\gamma)$ is negligible since $G(m_{\nu_S}^2/m_W^2)\to0$ for $m_{\nu_S}\ll m_W$.  

Finally, for clearness, we show in Table \ref{table8} an overview
of the main results of all considered models. Specifically,
we show energy scales for the neutrino masses and the ALPs parameter
space for each ISS case. 
\begin{table*}
\setlength{\tabcolsep}{0.5cm} \global\long\def\arraystretch{1.3}
\begin{centering}
\begin{tabular}{l c c c c l}
\hline 
\hline
\ \ ISS  &  $m_{a}$  & $g_{a\gamma}$  & \ $M$ (TeV),\ $\mu_{S}$ (keV) \tabularnewline
Model    & $\times10^{-11}$ eV  & $\times10^{-17}$ GeV$^{-1}$   & \hspace{0.2cm}$m_{\nu\textrm{light}}$ (eV)\tabularnewline
\hline 
\multirow{2}{1cm}{M22a}  & \multirow{2}{2cm}{$(19.0-56.0)$}  & \multirow{2}{1.8cm}{$(5.5-7.8)$}  & $(1.5-4.5),(0.1-0.7),$\tabularnewline
  &  &   & $(1.0-1.4)$\tabularnewline

\multirow{2}{0.8cm}{M22b}  & \multirow{2}{2cm}{$(0.52-1.4)$}  & \multirow{2}{1.8cm}{$(1.1-1.6)$}    & $(24.3-65.6),(11.3-58.9),$\tabularnewline
   &  &   & $(0.4-0.6)$\tabularnewline

\multirow{2}{0.8cm}{M33a}  & \multirow{2}{2cm}{$(19.0-56.0)$}  & \multirow{2}{1.8cm}{$(2.7-3.9)$}  & $(1.5-4.5),(0.1-0.7),$\tabularnewline
   &   &  & $(1.0-1.4)$\tabularnewline

\multirow{2}{0.8cm}{M33b}  & \multirow{2}{2cm}{$(0.52-1.4)$}  & \multirow{2}{1.8cm}{$(2.2-3.1)$}   & $(24.3-65.6),(11.3-58.9),$\tabularnewline
   &   &  & $(0.4-0.6)$\tabularnewline

\multirow{2}{0.8cm}{M23a} & \multirow{2}{3cm}{$(0.06-0.30)\times10^{11}$}  & \multirow{2}{2cm}{$(720-1200)$}  & $(7.5-21.0),(4.12-19.41),$\tabularnewline
   &    &  & $(1.33-2.24)$\tabularnewline

\multirow{2}{0.8cm}{M23b}  & \multirow{2}{3cm}{$(0.03-0.2)\times10^{11}$}  & \multirow{2}{2cm}{$(830-1400)$}   & $(5.6-15.8),(2.7-12.7),$\tabularnewline
   &    &  & $(1.4-2.5)$\tabularnewline
\hline 
\hline
\end{tabular}
\par\end{centering}
\protect\caption{Main features of the models discussed in the text. We have 
considered the constant $g\subset[10^{-3},2]$ in the mass term of
the ALPs, the $\eta$ Yukawa at order $10^{-2}$ in the M23a(b) models,  and $\Omega_{\textrm{DM}}^{\textrm{Planck}}h^{2}$ at $3\sigma$.
\label{table8}}
\end{table*}

\section{Discussion and summary \label{conclusion}} 
We have connected two interesting motivations for going beyond the standard model: neutrino masses and ALPs as dark matter. A natural scenario for achieving that is the ISS mechanism. In particular, we have considered the minimal versions of the ISS mechanism in agreement with all the neutrino constraints.  Nevertheless, in the considered framework, the mass scales for the ISS mechanism are generated by gravity-induced non-renormalizable operators when the scalar field containing the ALPs gets a vacuum expectation value, $v_{\sigma}$. Naturalness of these scales imposes strong constraints on these operators and, when combining these with the ALPs acceptable range for $v_{\sigma}$, only two solutions are possible: $(p,|q|) = (2,3)\ \ \text{for}\ \ 6\times10^{10}\lesssim v_{\sigma}\lesssim1\times10^{11}\ \textrm{GeV}$ and $(p,|q|)= (3,5)\ \ \text{for}\ \ 2\times10^{13}\lesssim v_{\sigma}\lesssim8\times10^{13}\ \textrm{GeV}$. This implies that operators given $M$ and $\mu_S$ scales can only belong to these two categories. Then, a simultaneous application of constraints coming from the texture of ISS mass matrix, the violation of the unitarity, the mass of exotic charged leptons, the stability of the effective Lagrangian against gravitational effects and the suitable ALPs parameter space ($m_a$ and $g_{a\gamma}$) to provide the total DM density almost set the rest of terms in the Lagrangian, only leaving a few of possibilities for all of ISS cases.
These constraints ultimatelly lead to a concrete prediction for the viable ALPs masses and ALPs-photon couplings and also for the mass scale of the heavy neutrinos necessary to explain the neutrino oscillation data. In other words, both sectors are deeply connected and the observation of a hypothetical signal of the ALP existence within the proper regions will automatically lead to the existence of heavy neutrino states in the TeV and multi TeV scales. In the same way, the nonobservation of an ALP within such regions or the observation of heavy neutrinos below the TeV scale would disfavour the possible linkage between ALP DM and neutrino masses suggested in this work.

Among the minimal ISS mechanisms, the $(2,2)$ and $(3,3)$ ISS cases are quite similar. It is due to the fact that in both of them $n_{N_{R}}=n_{S_{R}}$ implying that neutrino mass spectrum is characterized by only two mass scales, $M$ and $m_{\nu\text{light}}$. Thus, the results obtained are almost identical. Although, there is a slightly difference in the value of $g_{a\gamma}$ due to the presence of more fermions in the $(3,3)$ ISS case. In both cases, we find two effective models denoted as M22a,b and M33a,b in Table \ref{table8}. Since there is not sterile neutrino in these cases, the total DM density is made of ALPs.  We also remark that, although, the ALPs in these models can decay to two photons and, in the $(2,2)$ ISS mechanism, to two massless active neutrinos,  these are cosmologically stable because those decays are strongly suppressed by factors of $1/M_{\textrm{Pl}}^{2}$ and/or $1/v_{\sigma}^2$. 

On the other hand, the $(2,3)$ ISS case is phenomenologically more interesting due to the presence of a sterile neutrino in the mass spectrum. It implies that the DM density can be made of ALPs and $\nu_S$.  We have found a model satisfying all of previously mentioned constraints and, at the same time, offering the total DM. Because sterile neutrinos in the $(2,3)$ ISS mechanism can give, roughly speaking, at most $\approx43\%$ of the DM density, it is necessary that the remaining $\approx57\%$ of DM be made of ALPs. It is also possible that ALPs give the total DM density. It occurs when the mixing angle between active and sterile neutrinos is very suppressed in order to make the the Dodelson-Widrow mechanism inefficient. Both cases were studied in detail and denoted as M23a and M23b, respectively.

Regarding the search for ALPs, the benchmark regions in Figure \ref{fig1} are out of reach of the current and future experimental searches for axion/ALPs such as ALPS II, IAXO, CAST \cite{Graham:2015ouw}, since these currently have not enough sensitivity to probe the ALPs/axion-photon couplings and masses that are motivated in models with scales $v_\sigma\gtrsim 10^{13}$ GeV. 
Nevertheless, for the (2,2) and (3,3) ISS cases the benchmark regions are remarkably within the target regions in proposed experiments based on LC circuits \cite{Kahn:2016aff,Sikivie:2013laa}, which are designed to search for QCD axions and ALPs and cover many orders of magnitude in the parameter space of these particles, beyond the current astrophysical and laboratory limits \cite{Budker:2013hfa,Graham:2015ouw,Graham:2013gfa}. Specifically, the ABRACADABRA experiment \cite{Kahn:2016aff}  may explore ALPs masses as low as $\sim10^{-10}$ eV for a coupling to photons of the order of  $\sim10^{-18}$ GeV$^{-1}$, which are well below our benchmark regions (Figure \ref{fig1}).

Finally, despite the fact that neutrino mass spectrum is not completely predicted in the models found, the matrix scales in the ISS mechanism are estimated to be in agreement with the neutrino constraints \footnote{It is worth mentioning that for all the models the normal spectrum is the preferred neutrino mass spectrum \cite{Abada:2014vea} which in turn implies that our scan results are also compatible with the cosmological upper bound on the neutrino mass sum \cite{Hannestad:2016fog,Vagnozzi:2017ovm}}. Moreover, we have numerically checked, in all models, that there are solutions with coupling constants of order one that also satisfy LFV processes and the unitary condition. These processes can easily avoid without fine-tuning in the models discussed in this paper. Specifically, we have found that the BR($\mu\rightarrow e\gamma)$ in all cases are as small as $\sim10^{-20}-10^{-15}$ which are consistent with the current experimental value BR($\mu\rightarrow e\gamma) < 5.7\times10^{-13}$ \cite{Adam:2013mnn} and with future sensitivities around $6\times10^{-14}$ \cite{Baldini:2013ke}.

\begin{acknowledgments}
B. L. S. V. would like to thank Coordena\c{c}\~ao de Aperfei\c{c}oamento de
Pessoal de N\'ivel Superior (CAPES), Brazil, for financial support. C. D. R. C. acknowledges
the financial support given for the Departamento Administrativo de Ciencia, Tecnología e Innovación - COLCIENCIAS (doctoral scholarship 727-2015), Colombia, and the hospitality of Laboratori Nazionali di Frascati, Italy, in the final stage of this work.
O. Z. has been partly supported by UdeA/CODI grant IN650CE and by COLCIENCIAS through the Grant No. 111-565-84269.
\end{acknowledgments}

\bibliography{references}

\end{document}